\documentclass[useAMS,usenatbib]{mn2e}
\usepackage{graphicx,amsmath}
\usepackage{amssymb}
\usepackage{natbib}
\usepackage{threeparttable}
\usepackage{microtype}
\usepackage{color}
\usepackage{subcaption}
\def\CIVdblt{{\rm C~}\kern 0.1em{\sc iv}~$\lambda\lambda 1548, 1550$}
\def\MgIIdblt{{\rm Mg~}\kern 0.1em{\sc ii}~$\lambda\lambda 2796, 2803$}
\def\NVdblt{{\rm N}\kern 0.1em{\sc v}~$\lambda\lambda 1238, 1242$}
\def\OVIdblt{{\rm O}\kern 0.1em{\sc vi}~$ 1031, 1037$}
\def\SiIVdblt{{\rm Si~}\kern 0.1em{\sc iv}~$\lambda\lambda1393, 1402$}
\def\AlIIIdblt{{\rm Al~}\kern 0.1em{\sc iii}~$\lambda\lambda1855,1863$}
\def\FeIIdblt{{\rm Fe~}\kern 0.1em{\sc ii}~$\lambda\lambda 2383, 2600$}
\def\NeVIIIdblt{{\rm Ne~}\kern 0.1em{\sc viii}~$ 770, 780$}

\def\NeVIII{\hbox{{\rm Ne~}\kern 0.1em{\sc viii}}}
\def\OI{\hbox{{\rm O~}\kern 0.1em{\sc i}}}
\def\OII{\hbox{{\rm O~}\kern 0.1em{\sc ii}}}
\def\OIII{\hbox{{\rm O~}\kern 0.1em{\sc iii}}}
\def\OIV{\hbox{{\rm O~}\kern 0.1em{\sc iv}}}
\def\OV{\hbox{{\rm O~}\kern 0.1em{\sc v}}}
\def\OVI{\hbox{{\rm O~}\kern 0.1em{\sc vi}}}
\def\OVII{\hbox{{\rm O~}\kern 0.1em{\sc vii}}}
\def\OVIII{\hbox{{\rm O~}\kern 0.1em{\sc viii}}}
\def\NIII{\hbox{{\rm N~}\kern 0.1em{\sc iii}}}
\def\NIV{\hbox{{\rm N~}\kern 0.1em{\sc iv}}}
\def\NVII{\hbox{{\rm N~}\kern 0.1em{\sc vii}}}
\def\CIII{\hbox{{\rm C~}\kern 0.1em{\sc iii}}}
\def\SiIII{\hbox{{\rm Si~}\kern 0.1em{\sc iii}}}
\def\SVI{\hbox{{\rm S~}\kern 0.1em{\sc vi}}}
\def\NeIX{\hbox{{\rm Ne~}\kern 0.1em{\sc ix}}}

\def\AlII{\hbox{{\rm Al~}\kern 0.1em{\sc ii}}}
\def\AlIII{\hbox{{\rm Al~}\kern 0.1em{\sc iii}}}
\def\CaI{\hbox{{\rm Ca}\kern 0.1em{\sc i}}}
\def\CaII{\hbox{{\rm Ca}\kern 0.1em{\sc ii}}}
\def\CrII{\hbox{{\rm Cr}\kern 0.1em{\sc ii}}}
\def\CII{\hbox{{\rm C~}\kern 0.1em{\sc ii}}}
\def\CIII{\hbox{{\rm C~}\kern 0.1em{\sc iii}}}
\def\CIV{\hbox{{\rm C~}\kern 0.1em{\sc iv}}}
\def\CV{\hbox{{\rm C}\kern 0.1em{\sc v}}}
\def\H{\hbox{{\rm H}}}
\def\HI{\hbox{{\rm H~}\kern 0.1em{\sc i}}}
\def\HII{\hbox{{\rm H~}\kern 0.1em{\sc ii}}}
\def\Lya{\hbox{{\rm Ly}\kern 0.1em$\alpha$}}
\def\Lyb{\hbox{{\rm Ly}\kern 0.1em$\beta$}}
\def\Lyg{\hbox{{\rm Ly}\kern 0.1em$\gamma$}}
\def\Lyth{\hbox{{\rm Ly}\kern 0.1em$\theta$}}
\def\Lyfive{\hbox{{\rm Ly}\kern 0.1em$5$}}
\def\Lysix{\hbox{{\rm Ly}\kern 0.1em$6$}}
\def\Lyseven{\hbox{{\rm Ly}\kern 0.1em$7$}}
\def\Lyeight{\hbox{{\rm Ly}\kern 0.1em$8$}}
\def\Lynine{\hbox{{\rm Ly}\kern 0.1em$9$}}
\def\Lyten{\hbox{{\rm Ly}\kern 0.1em$10$}}
\def\HeI{\hbox{{\rm He}\kern 0.1em{\sc i}}}
\def\HeII{\hbox{{\rm He}\kern 0.1em{\sc ii}}}
\def\FeI{\hbox{{\rm Fe~}\kern 0.1em{\sc i}}}
\def\FeII{\hbox{{\rm Fe~}\kern 0.1em{\sc ii}}}
\def\FeIII{\hbox{{\rm Fe~}\kern 0.1em{\sc iii}}}
\def\MnII{\hbox{{\rm Mn}\kern 0.1em{\sc ii}}}
\def\MgI{\hbox{{\rm Mg~}\kern 0.1em{\sc i}}}
\def\MgII{\hbox{{\rm Mg~}\kern 0.1em{\sc ii}}}
\def\MgIII{\hbox{{\rm Mg~}\kern 0.1em{\sc iii}}}
\def\MgIV{\hbox{{\rm Mg~}\kern 0.1em{\sc iv}}}
\def\MgX{\hbox{{\rm Mg~}\kern 0.1em{\sc x}}}
\def\NaI{\hbox{{\rm Na}\kern 0.1em{\sc i}}}
\def\NV{\hbox{{\rm N}\kern 0.1em{\sc v}}}
\def\NII{\hbox{{\rm N}\kern 0.1em{\sc ii}}}
\def\NIII{\hbox{{\rm N}\kern 0.1em{\sc iii}}}
\def\OVI{\hbox{{\rm O}\kern 0.1em{\sc vi}}}
\def\SiII{\hbox{{\rm Si~}\kern 0.1em{\sc ii}}}
\def\SiIII{\hbox{{\rm Si~}\kern 0.1em{\sc iii}}}
\def\SiIV{\hbox{{\rm Si~}\kern 0.1em{\sc iv}}}
\def\SII{\hbox{{\rm S}\kern 0.1em{\sc ii}}}
\def\SIII{\hbox{{\rm S}\kern 0.1em{\sc iii}}}
\def\SIV{\hbox{{\rm S}\kern 0.1em{\sc iv}}}
\def\TiII{\hbox{{\rm Ti}\kern 0.1em{\sc ii}}}
\def\ZnII{\hbox{{\rm Zn}\kern 0.1em{\sc ii}}}
\def\kms{\hbox{km~s$^{-1}$}}
\def\cmsq{\hbox{cm$^{-2}$}}
\def\cc{\hbox{cm$^{-3}$}}
\def\etal{et~al.\ }
\definecolor{red}{rgb}{1.0,0,0}
\newcommand {\apgt} {\ {\raise-.5ex\hbox{$\buildrel>\over\sim$}}\ }
\newcommand {\aplt} {\ {\raise-.5ex\hbox{$\buildrel<\over\sim$}}\ } 

\title[{\OVI} - BLA Absobrer]{Detection of Low Metallicity Warm Plasma in a Galaxy Overdensity Environment at $\MakeLowercase{z} \sim 0.2$}

\author[Narayanan {\etal}]
{
\parbox{\textwidth}{ 
Anand Narayanan$^{2}$\thanks{Email: anand@iist.ac.in},
Blair D. Savage$^{3}$,
Preetish K. Mishra$^{4}$,
Bart P. Wakker$^{3}$,
Vikram Khaire$^{4,5}$,
Yogesh Wadadekar$^{4}$,
} 
\vspace*{10pt}\\ 
$^{1}$Based on observations with the NASA/ESA {\it Hubble Space Telescope}, obtained at the Space Telescope Science \\ Institute, which is operated by the Association of Universities for Research in Astronomy, Inc., under NASA contract NAS 05-26555\\
$^{2}$Indian Institute of Space Science \& Technology, Thiruvananthapuram 695547, Kerala, INDIA\\  
$^{3}$Department of Astronomy, The University of Wisconsin-Madison, 5534 Sterling Hall, 475 N. Charter Street, Madison WI 53706-1582, USA\\ 
$^{4}$National Centre for Radio Astrophysics, Tata Institute of Fundamental Research, Post Bag 3, Pune - 411007, INDIA.\\ 
$^{5}$Department of Physics, University of California Santa Barbara, CA, 93106, USA.   
}   

\begin{document}

\date{}

\pagerange{\pageref{firstpage}--\pageref{lastpage}} \pubyear{2017}
\maketitle

\label{firstpage}

\vspace{20 mm}

\begin{abstract}

We present results from the analysis of a multiphase {\OVI} - broad {\Lya} absorber at $z = 0.19236$ in the $HST$/COS spectrum of PG~$1121+422$. The low and intermediate ionization metal lines in this absorber have a single narrow component, whereas the {\Lya} has a possible broad component with $b(\HI) \sim 71$~{\kms}. Ionization models favor the low and intermediate ions coming from a $T \sim 8,500$~K, moderately dense ($n_{\H} \sim 10^{-3}$~{\cc}) photoionized gas with near solar metallicities. The weak {\OVI} requires a separate gas phase that is collisionally ionized. The {\OVI} coupled with BLA suggests $T \sim 3.2 \times 10^5$~K, with significantly lower metal abundance and $\sim 1.8$ orders of magnitude higher total hydrogen column density compared to the photoionized phase. SDSS shows 12 luminous ($> L^*$) galaxies in the  $\rho \leq 5$~Mpc, $|\Delta v| \leq 800$~{\kms} region surrounding the absorber, with the absorber outside the virial bounds of the nearest galaxy. The warm phase of this absorber is consistent with being transition temperature plasma either at the interface regions between the hot intragroup gas and cooler photoionized clouds within the group, or associated with high velocity gas in the halo of a $\lesssim L^*$ galaxy. The absorber highlights the advantage of {\OVI}-BLA absorbers as ionization model independent probes of warm baryon reserves. 

\end{abstract}
\begin{keywords}
galaxies: halos, intergalactic medium, quasars: absorption lines, quasars: individual:PG~$1121+422$, ultraviolet: general
\end{keywords}

\section{Introduction}\label{sec1}

The diffuse circumgalactic and intergalactic medium is predicted and observed to contain the vast majority of the cosmic budget of baryons compared to those that have condensed to form stars and the interstellar medium in galaxies \citep{persic92,rauch97,weinberg97,fukugita98,cen99,tripp00,cen06,bregman07,tepper11,tumlinson11b,savage14,wakker15,danforth16}. A considerable fraction ($\sim 40$\%) of these baryons are at shock-heated warm-hot temperatures of $T \sim 10^5 - 10^7$~K in the form of intergalactic filaments and sheets of the Cosmic Web, as intra-cluster and intra-group gas, and also in galaxy halos, as far out as several hundred kiloparsec. Examining this gaseous phase is of central importance to understanding how baryonic matter is reorganized in the universe as a result of structure formation, and for learning about the influence of the large-scale environments on the chemical and physical properties of gas outside of galaxies. 

A promising approach for probing this high temperature phase has been the use of absorption lines from ions such as {\OVI}, {\NeVIII}, and {\MgX}. However, the intergalactic (IGM) and the circumgalactic medium (CGM) are complex multiphase environments where the warm-hot medium is often kinematically entwined with high column density low ionization gas \citep[e.g.,][]{masiero05,savage10,savage11a,tumlinson11a,narayanan12,pachat16}. Careful analysis of individual absorption systems is necessary to sort out the physical conditions, chemical abundances, and baryonic columns in the low and high ionization gas phases of such absorbers and establish the presence of a warm-hot medium.

The cooler phase in multiphase absorbers is best studied through low ionization potential species such as {\CII}, {\OII}, {\SiII} and {\MgII}, with the intermediate ions such as {\CIII} and {\SiIII} adding additional valuable constraints for this phase. The low ionization medium is typically found to be of moderately high density ($n_{\H} \lesssim 10^{-3}$~{\cc}), with relatively low temperatures ($T \lesssim 3 \times 10^4$~K) and compact sizes $L \lesssim 100$~pc \citep[e.g.,][]{rigby02,tripp02,ding03a,ding03b,masiero05}. The ionization in this phase is largely regulated by the extragalactic UV background radiation. 

The physical conditions in the high ionization gas in multiphase absorbers, seen through the absorption signatures of {\OVI} ({\CIV}, and occasionally {\NV}) have often been ambiguous. Under collisional ionization equilibrium (CIE), the {\CIV} ionization fraction peaks at $T = 10^5$~K, and {\OVI} at $T = 2.5 \times 10^5$~K \citep{gnat07}. Both ions have strong resonant doublet lines in the UV, which make them suitable probes of warm gas. However, these high ions can also be produced in photoionized gas with low densities of $n_{\H} \sim 10^{-4} - 10^{-5}$~{\cc} and temperatures that are at least an order of magnitude lower compared to CIE \citep[e.g.,][]{tripp08, hussain17}. The ambiguity in ionization is a concern, as the metal abundances, absorber sizes and the total baryonic masses inferred tend to be completely different for photoionization and collisionally ionization models. The possibility of non-equilibrium ionization fractions due to delayed ion-electron recombination in a rapidly cooling gas has to be also considered while modeling the high ionization gas \citep{gnat07}. In some samples of multiphase absorbers, the presence of $T > 10^5$~K gas has become apparent through the detection of {\NeVIII} and {\MgX}, as the origin of such ions are best explained through collisional ionization \citep{savage05,narayanan09,narayanan11,savage11a,savage11b,tripp11,meiring13,hussain15,qu16,bordoloi16, pachat17}.

Theoretical models also yield divergent views on the origin of intervening {\OVI} absorbers. The numerical simulations of \citet{kang05, oppenheimer09} and \citet{oppenheimer12} lead to a photoionization explanation for {\OVI} absorbers, whereas \citet{smith11,tepper11} and \citet{mcquinn17} find {\OVI} is predominantly produced in warm and hot plasma, indicating that the ion cannot be directly used as a tracer of $T \gtrsim 10^5$~K shocked gas. 

The key to establishing the presence of a warm-hot medium in multiphase absorbers is to have ionization independent measurements of gas phase temperatures. In gas where the velocity dispersion is predominantly thermal, the different line widths of {\HI} and the metal lines can be used to establish the temperature. In some {\OVI} absorbers, the presence of a thermally broad {\Lya} (BLA, $b > 40$~{\kms}) component has been observed. Under pure thermal broadening, $b(\HI) > 40${\kms} corresponds to $T > 10^5$~K. The high $S/N$, low-$z$ {\OVI} survey of \citet{savage14} contains a list of 14 such absorbers (see their Table 5) for which the narrow and broad line widths of {\OVI} and {\HI} directly indicates the presence of gas with $T \gtrsim 10^5$~K. 

The challenge in this approach is in identifying the BLA feature amidst the strong absorptions from the cooler gas phases within the absorber. With the neutral hydrogen fraction being very low ($f_{\HI} = N(\HI)/N(\H) < 1.8 \times 10^{-5}$) at $T \geq 10^5$~K, the BLA is most likely a shallow feature that can go undetected if the $S/N$ at the {\Lya} absorption is poor, or if the absorption from the cold {\HI} gas is too strong and kinematically wide. In certain cases, the BLA materializes as broad wings on the narrow low ionization {\Lya} feature. The Cosmic Origins Spectrograph (COS), with its superior sensitivity over current and previous UV spectrographs, has enabled the discovery of many such BLA features in multiphae {\OVI} absorbers observed at high $S/N$ \citep[e.g.,][]{narayanan10b, savage11a, savage11b, savage12, savage14}.  

In this paper we report on the detection and analysis of a weak {\OVIdblt} doublet in a multi-phase intervening absorber. The strong {\Lya} absorption shows the likely presence of a BLA, which along with the {\OVI} points to the presence of a warm phase in the absorbing medium. In Sec. 2, we present information on the COS data. In Sec. 3, measurements on the multiphase absorber are described, along with a detailed analysis of the probable BLA detection. Sec. 4 deals with results from various ionization models for the absorber. In Sec. 5, we present information on galaxies in the vicinity of the absorber. We conclude in Sec 6 with a summary of the key results and a brief discussion on the possible astrophysical origin of this absorber. 

\section{Data Analysis}

The QSO PG~$1211+422$ \citep{Green86} has an emission redshift of $z_{em} = 0.225$ according to the reanalysis of SDSS quasar redshifts carried out by \citet{Hewett10}, which corrects
for systematic biases in the pipeline SDSS redshifts. The QSO was observed with COS \citep{Green12} in April 2011 as part of the instrument GTO program (Cycle 18, PID 12024, PI James Green). The spectra were obtained using the G130M and G160M gratings with integration times of 15.1ks and 17.4ks respectively. The separate exposures were retrieved from the $HST$/COS MAST archive\footnote{https://archive.stsci.edu/}, and processed using the STScI developed CalCOS (v3.0) pipeline. We used the method described by \citet{wakker15} to remove the wavelength offsets present in the pipeline processed individual exposures. The method involves cross-correlating the absorption lines in the individual spectra to establish the relative offsets as a function of wavelength. In the case of PG~$1121+422$, the offsets varied by up to $30$~{\kms} from one side of a segment to the other side. After applying the necessary shifts to produce a combined spectra, we aligned the Galactic ISM lines with the Galactic 21-cm emission. In spectral regions with no ISM lines, we used the lines of the Lyman series in the associated ($z = 0.21979$) absorber to determine the offsets. In the case of PG~$1121+422$, this alignment required a second order polynomial fit to the offset as a function of wavelength, with shifts varying from $-20$~{\kms} near $1200$~{\AA} to $+40$~{\kms} near $1750$~{\AA}. The final relative and absolute offsets were estimated as better than $5$~{\kms}. This procedure removes the issue that COS spectra that are coadded without adequately correcting for the wavelength dependent offsets is likely to produce blurred versions of spectral line features. 

The individual exposures thus adjusted were combined by adding the counts in each pixel, and converting the sum back to a flux. The errors were calculated using pure Poisson statistics, which has empirically been shown to produce the correct result when measuring the rms around a fitted continuum \citep{wakker15}. At the brightness level of PG~$1121+422$, the background and dark counts are negligible. 

Detector fixed pattern noise features are mitigated to a large extent in the coaddition process itself, as the exposures were obtained with different G130M and G160M grating central wavelength settings. The coadded spectrum was resampled to two wavelength pixels per resolution element (FWHM $ = 17 - 20$~{\kms}). The resultant spectrum with wavelength coverage from $1130 - 1800$~{\AA} has $S/N \sim 18,~20,~11$ per resolution element at the redshifted locations of {\Lya}, {\OVI}~$1031$ and {\SiIV}~$1393$ of the $z=0.19236$ absorber. The spectrum was normalized by fitting low-order polynomials to the continuum over wavelength intervals of 20~{\AA}.  

\section{Measurements on the Absorber at $z = 0.19236$}

In the $HST$/COS spectrum, the $z = 0.19236$ absorber is detected through its {\HI} Lyman lines, {\CII}, {\SiII}, {\CIII}, {\SiIII}, {\SiIV}, and {\OVI} lines. Additionally, COS offers coverage of transitions of {\NII}, {\NIII}, and {\NV}, which are non-detections at $\geq 3\sigma$ significance. In Figure 1, we show the various absorption lines covered by COS. The $HST$/$Faint~Object~Spectrograph$ (FOS) archival data for this sightline provide an upper limit on the {\CIVdblt} lines which are also non-detections. The equivalent width measurements on the various lines are given in Table 1. The rest-frame equivalent widths $W_r(\CII~1334) = 158~{\pm}~17$~m{\AA}, and $W_r(\SiII~1260) = 125~{\pm}~10$~m{\AA} indicate that this is a weak {\MgII} class of absorber \citep{Narayanan05}. In other words, the system would have $W_r (\MgII~2796) < 300$~{m\AA} if there was coverage of the {\MgIIdblt} doublet lines. Weak {\MgII} absorbers are optically thin in {\HI}, with unsaturated and kinematically simple metal line features \citep{rigby02}. As a separate class of metal line absorbers, they are thought to have a different astrophysical origin compared to strong {\MgII} systems with $W_r(\MgII~2796) \geq 300$~m{\AA}. The strong {\MgII} systems have kinematically broad ($\Delta v > 200$~{\kms}) and saturated metal lines whose incidence is well correlated with $L > 0.05L^*$ galaxies \citep{cwc05,kacprzak11a} at close impact parameters \citep[$\rho < 100$~kpc][]{chen10b,bordoloi11}. The absorber number density evolution with redshift is different for the strong and weak {\MgII} absorbers, an additional indication that the two class of absorbers are of different origin \citep{Narayanan05,nestor05,prochter06,lynch06,narayanan08,zhu13}.

\begin{figure*} 
\centerline{
\vbox{
\centerline{\hbox{ 
\includegraphics[angle=00,width=0.7\textwidth]{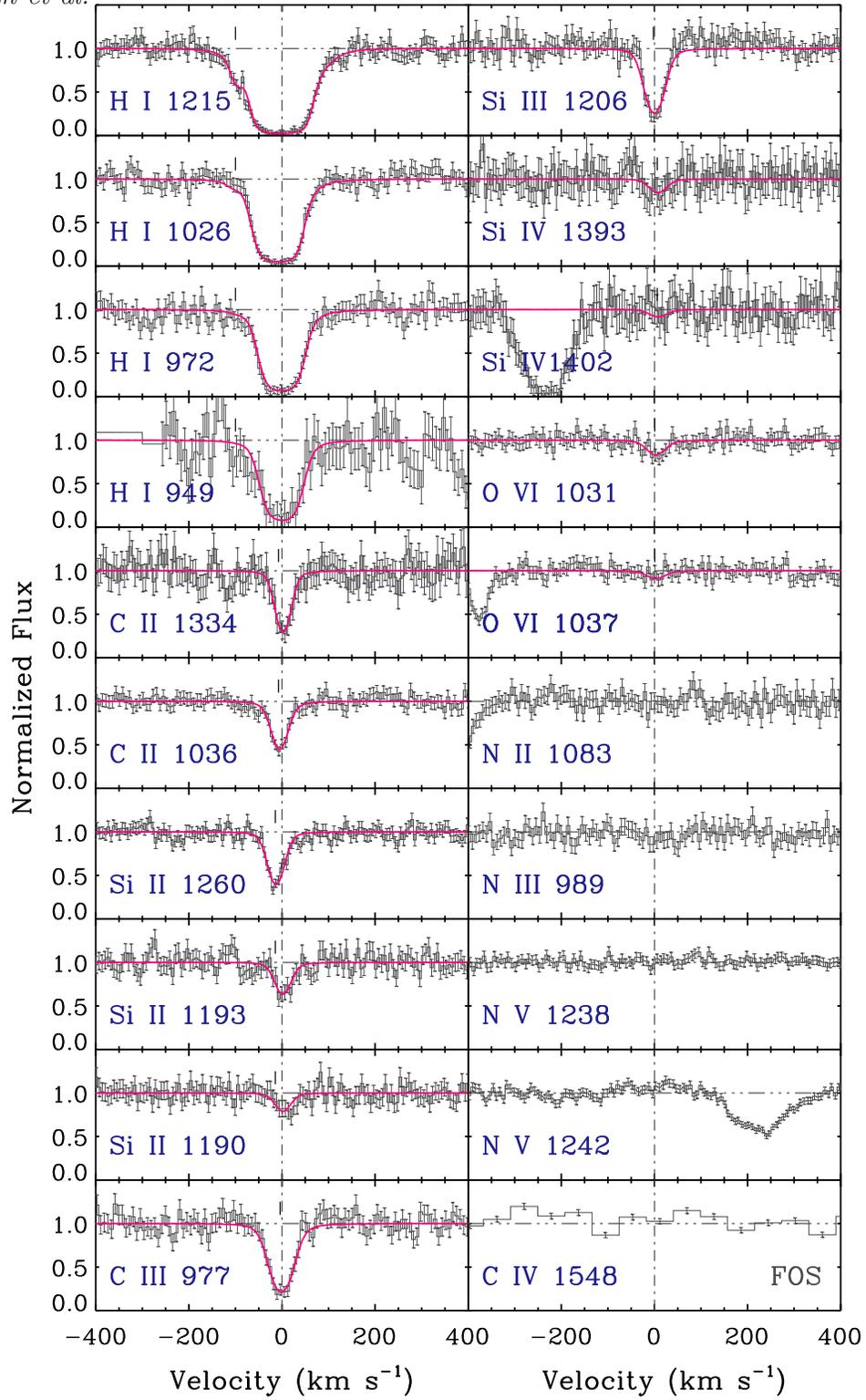} 
}}
}}  
\caption{Continuum normalized spectra of relevant transitions associated with the $z = 0.19236$ absorber towards PG~$1121+422$. The horizontal axis is velocity in the rest-frame of the absorber, where $v = 0$~{\kms} stands for $z = 0.19236$. The vertical error bars represent $1\sigma$ uncertainty in flux values. Overlaid are Voigt profile model fits whose parameters are given in Table 1. The {\CIV} information comes from $HST$/FOS spectra. The {\HI} lines require a two component model profile, whereas all the remaining lines are explained by a single component.}  
\label{fig:1} 
\end{figure*}       

We applied Voigt profile fits to the detected lines using the \citet{fitzpatrick97} routine. The model profiles were convolved with the empirically determined line-spread functions of COS by \citet{kriss11} at the observed wavelength of each line. Simultaneous line fitting was employed wherever more than one line from a single species was detected. The fit results are shown in Figure 1 and Table 2. In addition, line measurements done using the apparent optical depth (AOD) method of \citet{savage91} are listed in Table 1. The AOD method provides the true column density for unsaturated lines and is useful to reveal the presence of unresolved line saturation (a saturated line appearing unsaturated because of the instrumental broadening of the line feature). 

\begin{table*} 
\caption{AOD Measurements of Transitions associated with the $z=0.19236$ absorber} 
\begin{center}  
\begin{tabular}{lrrr}
\hline
Transition	&    $W_r$ (m\AA) &  $\log~[N_a,~{\cmsq}]$	   &    [$-v,~+v$]~(\kms) \\
\hline
\hline
{\HI}~$1215$	&	$706~{\pm}~11$	  &	$> 14.5$	   & 	[-175, 175] \\
{\HI}~$1026$	&	$393~{\pm}~14$	  &	$> 15.2$	   &	[-150, 150] \\
{\HI}~$972$	&	$356~{\pm}~12$	  &	$> 15.5$	   &	[-115, 115] \\
{\CII}~$1334$	&	$158~{\pm}~17$	  &	$14.05~{\pm}~0.06$ &	[-60, 60] \\
{\CII}~$1036$	&	$103~{\pm}~10$	  &	$14.06~{\pm}~0.05$ &	[-60, 60] \\
{\NII}~$1083$	&	$< 30$	  	  &	$< 13.4$ 	   &	[-60, 60] \\
{\SiII}~$1260$	&	$125~{\pm}~9$	  &	$13.00~{\pm}~0.05$ &	[-60, 60] \\
{\SiII}~$1193$	&	$78~{\pm}~11$	  &	$13.10~{\pm}~0.09$ &	[-60, 60] \\
{\SiII}~$1190$	&	$44~{\pm}~11$	  &	$13.12~{\pm}~0.09$ &	[-60, 60] \\
{\CIII}~$977$	&	$162~{\pm}~9$	  &	$13.63~{\pm}~0.05$ &	[-60, 60] \\
{\NIII}~$989$	&	$< 27$	  	  &	$13.5$ 		   &	[-60, 60] \\
{\SiIII}~$1206$	&	$156~{\pm}~9$	  &	$13.09~{\pm}~0.05$ &	[-60, 60] \\
{\SiIV}~$1393$	&	$33~{\pm}~20$	  &	$12.68~{\pm}~0.29$ &	[-60, 60] \\
{\SiIV}~$1402$	&	$< 63$	  	  &	$< 13.2$ 	   &	[-60, 60] \\
{\NV}~$1238$	&	$< 22$	  	  &	$< 13.0$ 	   &	[-60, 60] \\
{\OVI}~$1031$	&	$33~{\pm}~9$	  &	$13.48~{\pm}~0.12$ &	[-60, 60] \\
{\OVI}~$1038$	&	$< 24$	  	  &	$< 13.6$ 	   &	[-60, 60] \\
{\CIV}~$1548$ (FOS) &   $< 72$		  &     $< 13.2$	   &    [-60, 60] \\
\hline
\hline
\end{tabular}
\label{tab:tab1}
\end{center}
\scriptsize{The different columns are the equivalent width of the transition in the rest-frame of the absorber, the integrated apparent column density, and the velocity range of integration. For strongly saturated lines, a lower limit is quoted for the column density, and for lines not detected at $\geq 3\sigma$, an upper limit is quoted. The {\SiIV}~$1393$ line is formally a $1.7\sigma$ detection because of the low $S/N$ in the redshifted region of the spectrum. Nonetheless, the presence of a distinct absorption feature at the expected location (see Figure 1) is why we have listed a formal measurement for this transition.}
\end{table*}

The {\OVI} absorption is weak, and is detected at $\geq 3\sigma$ only in the $1031$~{\AA} line of the doublet. The absorber rest-frame equivalent width of $W_r(\OVI~1031) = 33~{\pm}~9$~m{\AA} we measure for this line is consistent with the non-detection of the {\OVI}~$1037$ transition at $\geq 3\sigma$. We note the presence of a weaker feature at the anticipated location of {\OVI}~$1037$ ($\lambda = 1237.24$~{\AA}), with $W_r = 12~{\pm}~7$~m{\AA} and a total column density of $\log~[N_a,~{\cmsq}] = 13.29~{\pm}~0.39$ obtained by integrating the apparent column densities, $N_a(v)$, over the velocity interval [$-65, +65$]~{\kms}. The $N_a(v)$ comparison of Figure 2 shows this feature agreeing with the {\OVI}~$1031$ over [$-40, +40]$~{\kms}, the interval within which the $1031$~{\AA} absorption is clearly evident. We infer the feature to be {\OVI}~$1037$ with a detection significance of $1.7\sigma$. Its presence, albeit at low significance, augments the detection of {\OVI}~$1031$, with a joint significance of $4.0\sigma$ for the {\OVIdblt} doublet (assuming that the equivalent width errors are dominated by photon count statistics). We also considered the prospect of {\OVI}~$1031$ being a weak {\Lya} feature at $z = 0.01216$. The {\Lya} has a redshift path length of $\Delta z = 0.012$ over this particular sightline, considering the path length up to the redshifted wavelength of {\OVI}~$1031$. The $dN(\Lya)/dz \sim 178$ at low-$z$ for weak lines with $13.1 < \log~N(\HI) < 14.0$ (Williger {\etal}2006) does not strongly favor ($N \sim 2$) the feature being an interloping {\Lya}. The line identification for this sightline does not associate the feature at the redshifted wavelength ($\lambda = 1230.46$~{\AA}) of {\OVI}~$1031$ with metal lines from any of the other absorbers along this sight line. The line list published by \citet{danforth16} also agrees with this\footnote{https://archive.stsci.edu/prepds/igm/}. Thus, the possibility of the {\OVI}~$1031$ being an interloping feature is small. 

\begin{table*} 
\caption{Voigt Profile Fit Measurements of Transitions associated with the $z=0.19236$ absorber} 
\begin{center}  
\begin{tabular}{lrrr}
\hline
Transition     &     $v$ (\kms)    	   &       log~$[N~(\cmsq)]$	&	$b$(\kms)     \\
\hline 
\hline 
{\HI}~$1215 - 972$  & $-101~{\pm}~4$  &   $12.97{\pm}~0.10$   &   $11~{\pm}~3$     \\
	 	    & $-4~{\pm}~4$    &   $16.44~{\pm}~0.09$  &   $25~{\pm}~4$     \\
		    & $1^a$	      &   $14.02~{\pm}~0.14$  &   $71~{\pm}~4^b$     \\	

\\             
{\OVI}~$1031, 1037$  & $1{\pm}~5$    &   $13.46~{\pm}~0.09$  &   $21~{\pm}~3$    \\
\\
{\SiII}~$1260, 1193, 1190$ & $-2~{\pm}~3$ & $13.10~{\pm}~0.11$	&  $18~{\pm}~4$ \\
\\
{\CII}~$1334,1036$  &  $1~{\pm}~3$ & $14.12~{\pm}~0.06$	&  $17~{\pm}~4$ \\
\\
{\CIII}~$977$	&      $-4~{\pm}~3$  & $13.84~{\pm}~0.10$	&  $23~{\pm}~3$ \\
\\
{\SiIII}~$1206$	&    $-2~{\pm}~2$    & $13.20~{\pm}~0.09$	&  $21~{\pm}~3$ \\
\\
{\SiIV}~$1393, 1402$  &  $6~{\pm}~5$  & $12.66~{\pm}~0.21$	&  $22~{\pm}~4$ \\
\\
\hline
\hline
\end{tabular}
\label{tab:tab2}
\end{center}
\scriptsize{Comments - (a) The third {\HI} component was obtained by fixing the centroid of the component at the same velocity as {\OVI}. (b) The BLA's $b$-parameter is poorly constrained. Taking into account systematic uncertainties (discussed in Sec 2), we adopt a value of $71^{+22}_{-14}$~{\kms}.}
\end{table*}

\begin{figure*}
\centerline{
\vbox{
\centerline{\hbox{ 
\includegraphics[angle=90,width=0.9\textwidth]{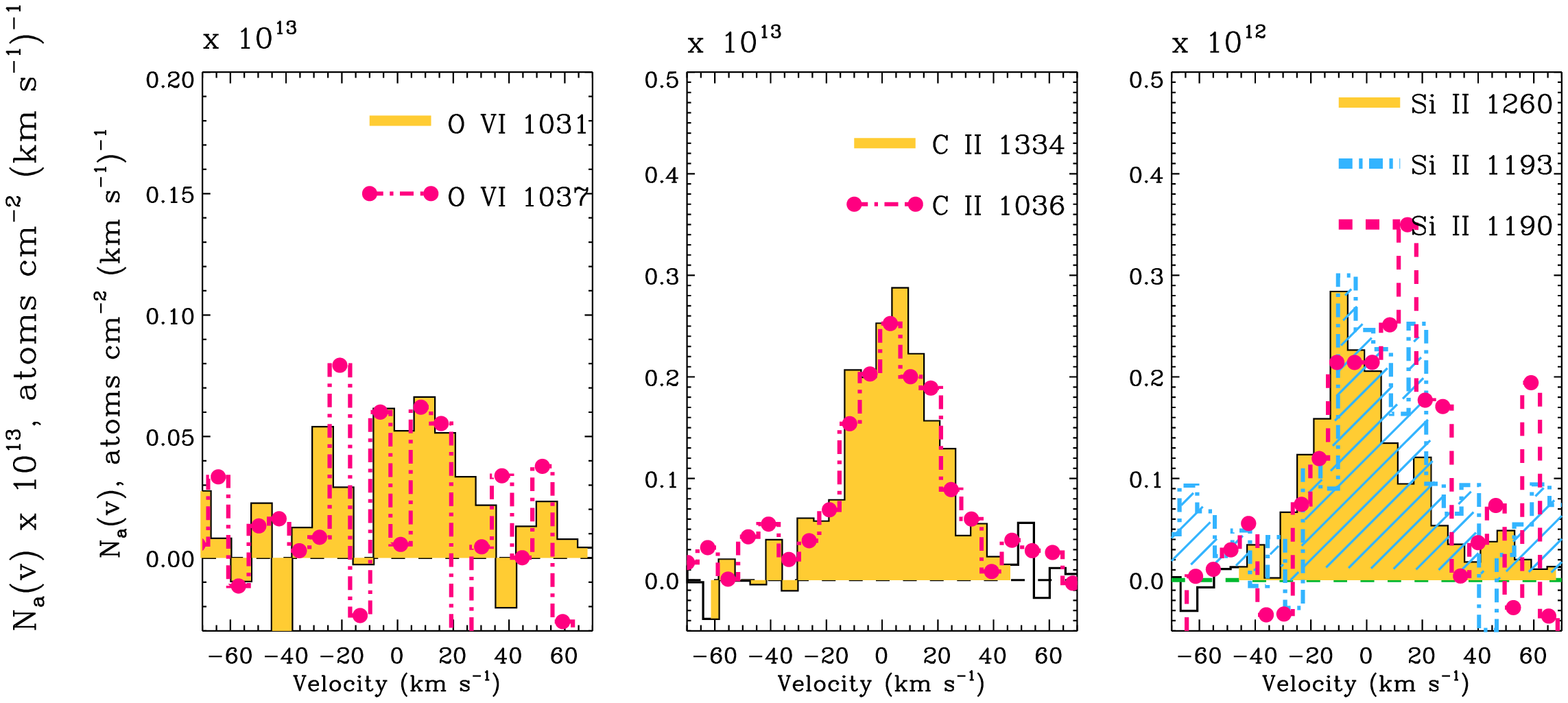} 
}}
}}  
\caption{The figure compares the run of apparent column densities of various transitions against system velocity. The quantity along the Y-axis has to be scaled with the mutliplicative factor given at the top of each panel. The \textit{left panel} shows {\OVI}~$1031$ and $1037$ lines. The $1037$ line is not a formal detection at $\geq 3\sigma$ significance, but a weak absorption ($W_r = 12~{\pm}~7$~m{\AA}) is seen at the expected location of the {\OVI}~$1037$ line. The \textit{middle panel} show both {\CII} transitions as consistent with each other with no compelling evidence for unresolved saturation or contamination. The \textit{right panel} show {\SiII} transitions. At the core of the absorption, the weaker {\SiII}~$1190$ line has slightly higher apparent column density compared to the two other lines of the multiplet, indicating that the stronger {\SiII} lines are midly saturated. We take this into saturation effect account while adopting a value from profile fitting for the column density of {\SiII} as explained in Sec 2. We also see mild excess absorption in the $1193$ line $v \sim +40$~{\kms} which could be contamination, as there is no evidence for such a component in the other two {\SiII} lines or the {\CII} lines.}  
\label{fig:2} 
\end{figure*} 

The {\HI} absorption is seen in Ly$\alpha - \delta$ transitions. The Ly$\delta$ line is at the extreme blue edge of the COS spectrum where $S/N$ is poor. Lyman lines of higher order are at wavelengths outside the coverage of COS. The Lyman transitions are strongly saturated. The {\Lya} line shows the presence of at least two kinematically separate components; a strong and heavily saturated core absorption at $v \sim 0$~{\kms}, and a narrower, weak absorption between $-140 \leq v \leq -80$~{\kms}. This second component is too shallow to be seen in the higher order Lyman lines. The fit with such a two component model is shown in Figure 3. The weak component at $v \sim -101$~{\kms} is possibly a little narrower and stronger than what the fitting routine recovers. The best fit is achieved when an additional broad component is added to explain the edges and wings of {\Lya}, especially redward of the core absorption. In Sec 3.1, we discuss the {\HI} profile fit results in greater detail. 

Using the apparent optical depth (AOD) method of \citet{savage91},  we measure column densities of {\CIII}~$977$ and {\SiIII}~$1206$ as $\log~N_a(\CIII) = 13.63~{\pm}~0.05$ and $\log~N_a(\SiIII) = 13.09~{\pm}~0.05$. These are $0.21$~dex and $0.11$~dex lower than the profile fit column densities for the {\CIII}~$977$ and {\SiIII}~$1206$ respectively, implying that these lines are saturated. AOD method underestimates the column density of unresolved lines. The profile fit results are more reliable, as they take line saturation into account to some extent. The Voigt profile models for these two lines are not unique. The fitting routine permits the $b$ and $N$ to vary over a small range of values, yielding acceptable fits to the line features. The true uncertainty in the line parameters derived from the fit should account for this. Keeping this in mind, we adopt a value of $\log~N(\CIII) = 13.84^{+0.38}_{-0.07}$ by varying the $b$ values in the range of $17 - 25$~{\kms}, outside of which the models do not fit the data well. The range of $b$ values is based on the line widths of the low ions ({\CII}, {\SiII}) and the core {\HI} which dominates the absorption. A similar analysis on {\SiIII}~$1206$ gives $\log~N(\SiIII) = 13.20^{+0.23}_{-0.08}$~{\kms}. Both {\CIII} and {\SiIII} lines closely follow the core absorption in {\HI}, suggestive of similar gas phase origins. 

The agreement between the $N_a(v)$ profiles of the multiple lines of {\CII} rules out strong saturation effects in those lines (see Figure 3). For {\SiII}, the weaker lines ($1190$~{\AA},~$1193$~{\AA}) show slightly higher $N_a(v)$ profiles than the stronger ones, with a difference of $0.12$~dex between integrated apparent column densities ($N_a$) of $1190$~{\AA} (weakest of the three transitions) and the $1260$ (strongest) lines. However, the individual AOD column densities of the {\SiII}~$1260,~1193,$ and $1190$ transitions are within the $1\sigma$ uncertainty in their column densities, and also from the simultaneous profile fitting of the three lines (see Table 1), suggesting only minor levels of unresolved saturation. Furthermore, applying the AOD method to data which is not of high $S/N$ is known to overestimate the column densities because of the inclusion of noisy pixels during apparent column density integration \citep{fox05}.  

\subsection{The Detection of Broad {\HI} Absorption}

A single component simultaneous fit to the Lyman transitions explains the strong absorption in the core of {\Lya} with some certainty. A second narrow component is needed to fit the weak absorption at $v \sim -101$~{\kms}. Such a two component fit, however, does not trace the absorption along the {\Lya}'s wings, particularly across the velocity interval of [$+65, +150$]~{\kms}. The disagreement between the model fit and the observed line is illustrated in the top panel of Figure 4. 

\begin{figure*}
\begin{center}
\centerline{
\vbox{
\centerline{\hbox{ 
\includegraphics[angle=90,width=0.8\textwidth]{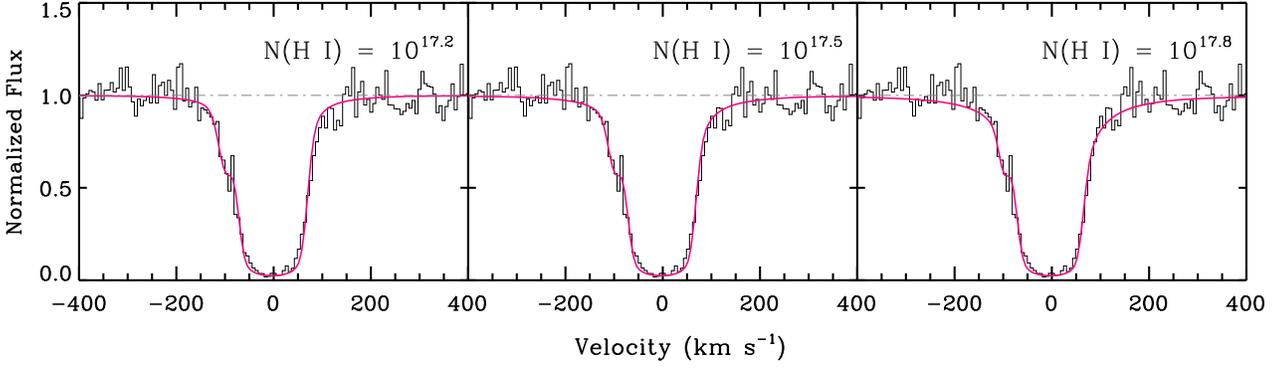} 
}}
}}  
\caption{The figure compares synthetic {\Lya} with Lyman limit column densities of $N(\HI) = 10^{17.2},~10^{17.5},~10^{17.8}$~{\cmsq} (\textit{solid red curve}) with the observed {\Lya} profile. Within this range, the lower values of {\HI} column density tend to result in larger $b$-values and broader profile fits in the core regions of the absorption. For $N(\HI) > 10^{17.5}$~{\cmsq}, the synthetic profiles are not consistent with {\Lya} or the higher order Lyman lines in the wings of the absorption profile, suggesting that the {\HI} in this absorber is possibly sub-Lyman limit.}  
\label{fig:2} 
\end{center}
\end{figure*} 

\begin{figure*}
\centerline{
\vbox{
\centerline{\hbox{ 
\includegraphics[angle=90,width=0.7\textwidth]{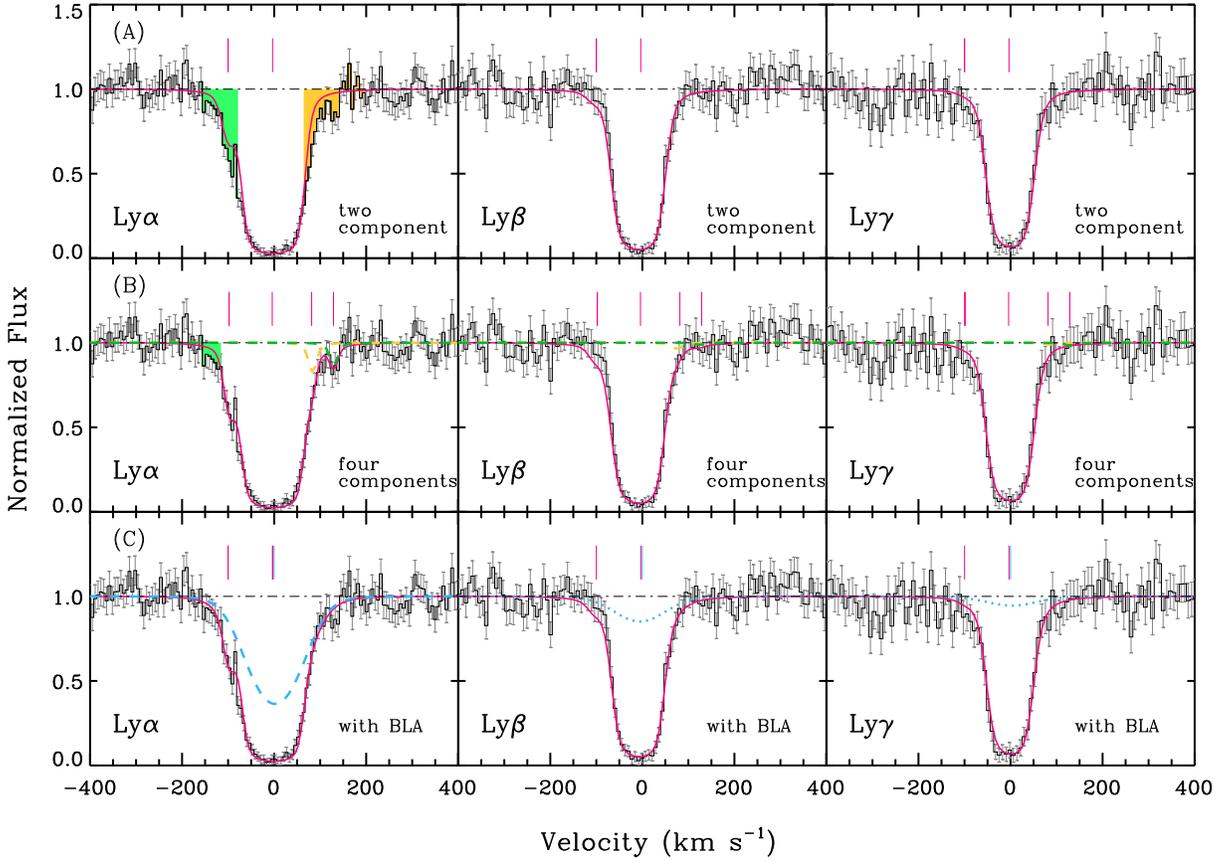} 
}}
}}  
\caption{The figure compares {\HI} profile fit models with and without a BLA. The \textit{top panel (A)} shows {\HI} lines modeled using a combination of two Voigt components at $v \sim -101$ and $v \sim -4$~{\kms} (\textit{red} curve). Such a model does not explain the absorption at the velocity edges of the {\Lya} profile (indicated by the shaded regions). In the \textit{middle panel (B)} we have attempted to explain the absorption in the positive velocity wing of the {\Lya} with a free fit involving two narrow components whose contributions are shown as \textit{yellow} dashed lines centered at $v = 82$ and $129$~{\kms}. The velocities for the components thus obtained are significantly offset from {\OVI} with $b(\HI) << b(\OVI)$. Such an approach of using multiple narrow components to explain the absorption along the positive and negative velocity wings of {\Lya} does not reveal the {\HI} that is associated with {\OVI}. In the \textit{bottom panel (C)} are profiles with a combination of the narrow components shown in panel (A) and a third component which is broad but shallow (dashed, \textit{blue} curve). The shallow component has a $b(\H) = 71$~{\kms} and is centered at the velocity of {\OVI}. The composite of these three components is the \textit{red} curve. The component centroids are indicated by the vertical tick marks. Statistically, such a three component model yields a slightly better fit ($\chi^2_{\nu} = 0.8$) than a two component fit ($\chi^2_{\nu} = 1.1$) of panel (A). The additional broad component helps to match the absorption in the positive and negative velocity wings of {\Lya} profile simultaneously. The fit values are given in Table 1.}  
\label{fig:3} 
\end{figure*}

As the {\HI} lines are strongly saturated, there exists the possibility of the column density being higher than the $\log~N(\HI) \sim 10^{16.44}$~{\cmsq}, the value profile fitting results yield. The lack of coverage of the redshifted Lyman limit for this absorber force us to resort to indirect means to establish the true total {\HI} column density. In Figure 3 we show synthetic profiles of {\Lya} for {\HI} column densities of $N \leq 10^{17.2}$~{\cmsq} (LLS, $\tau_{912\mathrm{{\AA}}} > 1$) overlaid on the data. The additional weak and narrow component at $v \sim -101$~{\kms} is also added to the synthetic profile. We find column densities in the range $10^{17.2} \lesssim N(\HI) \lesssim 10^{17.5}$~{\cmsq} for the core component, resulting in profiles with $b$-values of $25 - 21$~{\kms}, tracing the absorption in the wings of the {\Lya},  suggesting that the absorber could be a Lyman limit system. However, the model absorptions for $N(\HI) > 10^{17.2}$~{\cmsq} tend to be  a bit stronger than what the data suggests in the core regions of the profile ($-50 \lesssim \Delta v \lesssim 50$~{\kms}) as can be seen from Figure 3. An alternative to this is the possible presence of a third {\HI} component to explain the {\Lya} feature, in which the core {\HI} column density remains sub-Lyman limit. We prefer this latter scenario, among the two possible scenarios, for the following reasons. The redshift number density ($dN/dz$) of strong {\MgII} absorbers closely follows the $dN/dz$ of Lyman limit systems \citep{stengler-larrea95,nestor05,menard09} over $0.4 < z < 2$. Thus, the weak {\MgII} absorbers ought to be sub-Lyman limit systems with $10^{15.8} < \log~N(\HI) < 10^{16.8}$ \citep{cwc99,rigby02,Narayanan05}. The strength of {\CII} and {\SiII} lines in this absorber makes it a typical weak {\MgII} system \citep{Narayanan05}, whose accompanying {\HI} we expect to be sub-Lyman limit. In addition, the redshift path length for {\Lya} in this COS spectrum is $0 \leq z \leq 0.22$. From the $dN/dz \sim 0.25$ for LLS at $z < 0.5$ \citep{ribaudo11}, their anticipated incidence in the available path length comes out as low ($N \sim 0.06$), implying a low probability for the absorber to be optically thick at the Lyman limit. Lastly, ionization models (see Sec 4.1) imply the {\OVI} in this absorber traces a gas phase different from the low and intermediate ions. The hydrogen associated with this {\OVI} phase would be a separate component contributing to the {\HI} absorption. Based on these, we proceed with a three component solution to the {\HI} to account for the absorption in the core and along the wings.  

\begin{figure*}
\centerline{
\vbox{
\centerline{\hbox{ 
\includegraphics[angle=90,width=0.9\textwidth]{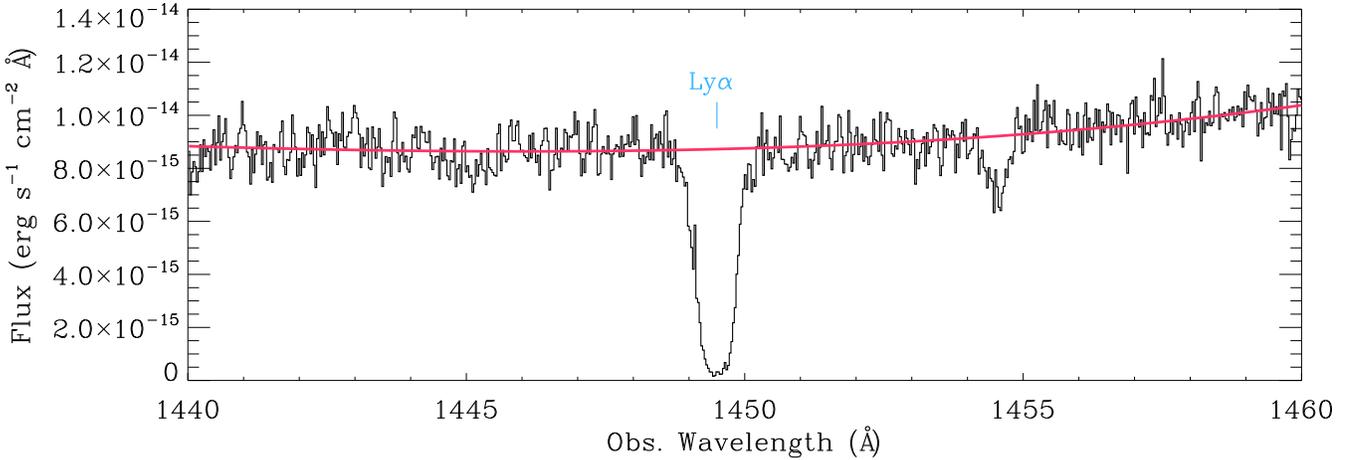} 
}}
}}  
\caption{The COS spectrum over a $20$~{\AA} region around the {\Lya} line associated with the $z=0.19236$ absorber. The smooth red curve is the continuum model for the region. The features at $1445.0$~{\AA} and $1454.8$~{\AA} are likely to be {\Lya} absorptions from $z=0.189$ and $z=0.197$ respectively.}  
\label{fig:4} 
\end{figure*} 

In Figure 4, we compare the results from the earlier two component model with a new three component Voigt profile model. A simultaneous fit to the Lyman transitions with three components result in the third component being kinematically broad and shallow. While fitting, the third component's velocity was fixed to the velocity of {\OVI}. Keeping the velocity as a free parameter does not allow the fitting routine to converge to a meaningful solution (large uncertainties in the derived $v$ and $b$ values) as this third component is shallow and is blended with the absorption in the core. The narrow component dominates the absorption in the core, whereas the broader component contributes more towards the absorption in the wings of the profile, especially between [$+65, +150$]~{\kms} of {\Lya}. Statistically, such a three component model offers an improved fit with reduced $\chi_{\nu}^2 = 0.8$ than a fit without the broad component ($\chi_{\nu}^2 = 1.1$). 

In the three component fit, the broad component with $b(\HI) \sim 71$~{\kms} is, by definition, a BLA. The $b$ value of the BLA has to be firmly established, as it is the basis for estimating the gas temperature in the absorber, independent of modeling assumptions. The formal errors from the fitting routine (listed in Table 1), ignore the few potential sources of systematic uncertainties, primarily the ambiguity in continuum placement and in the line parameters of the core {\HI} component. The continuum $20$~{\AA} along both sides of the {\Lya} is fairly featureless (see Figure 5), well defined with a polynomial of order 3, and hence likely to be of limited concern to the BLA fit results. 

To assess the core component's influence on the BLA properties, we lowered the $b$ paramter of the core absorption from $25$~{\kms} to $21$~{\kms}, below which it will not be realistic as the hydrogen would be narrower than the intermediate ionization metal lines. With $b(\HI) = 21$~{\kms}, the core absorption is fitted with $\log~N(\HI) = 17.18~{\pm}~0.09$, and the BLA with $b(\H) = 57~{\pm}~4$~{\kms} and $\log~N(\HI) = 14.30~{\pm}~0.16$. In this case, the {\HI} from the core, close to being a Lyman limit system, would begin to contribute towards the absorption in the wings as seen in Figure 3. Such a high column density for the core {\HI} is not preferred for reasons mentioned earlier in this section. Nonetheless, we use it for establishing limits on the BLA line parameters. With $b(\HI) = 29$~{\kms} for the core component, the BLA parameters come out as $b(\HI) = 93~{\pm}~5$~{\kms} and $\log~N(\HI) = 13.63~{\pm}~0.17$. Raising the $b$ of the core absorption in {\HI} to values greater than $29$~{\kms} result in profile models that are inconsistent with the {\Lyb} and {\Lyg} line shapes. Based on these, the BLA parameters are modified to $\log~N(\HI) = 14.02~{\pm}~0.17$ and $b(\H) = 71^{+22}_{-14}$~{\kms} accounting for the systematic uncertainties in the estimation of the core absorption in {\HI}. 

A possible alternative to a BLA is the presence of a few narrow components to explain the excess absorption in the interval [$+65, +150$]~{\kms} of {\Lya}. The number of components required to do this is uncertain as no well defined kinematic substructure is seen for {\Lya} at $v > 60$~{\kms}. We tried fitting the {\HI} by replacing the BLA with two narrow components, the results of which are shown in the middle panel of Figure 4. The best fit model places these two additional components at $v = +82~{\pm}~4$~{\kms} and $v = +129~{\pm}~3$~{\kms}, with column densities of $\log~N(\HI) = 12.71~{\pm}~0.40$, and $\log~N(\HI) = 12.81~{\pm}~0.51$, but very narrow $b$-values of $b(\HI) = 4~{\pm}~1$~{\kms} and $3~{\pm}~2$~{\kms} respectively, indicating that these components are much narrower than the resolution of COS. A similar narrow and weak component would be additionally required to explain the excess absorption between [$-180,-100$]~{\kms}. Such a model with three additional weak and narrow components in place of a broad-{\HI} component does not explain the {\HI} associated with {\OVI}. The additional components end up being too narrow for the relatively broad {\OVI}, as the profile fitting exercize for {\HI} reveals. In contrast, the excess absorption at both the positive and the negative velocity ends of the {\Lya} are simultaneously explained by a single BLA component (see bottom panel of Figure 4).  

\section{Ionization Modeling}

To determine the physical properties and chemical abundances in the absorber, we explored photoionization equilibrium models, pure collisional ionization equilibrium models, and hybrid models that consider both photoionization and collisional ionization equilibrium processes concurrently. The results from these are explained in the subsequent sections. In the models we use the \citet{ks15b} model of the extragalactic ionizing UV background for $z = 0.19$ as the ionizing radiation. The conventional \citet{hm12} ionizing background underestimates the {\HI} photoionization rate by a factor of $\sim 1.5 - 3$ at low-$z$ \citep{kollmeier14, shull15, wakker15, gaikwad17a, viel17}. \citet{ks15b} resolve this by adopting the recent estimates of QSO emissivity and star formation history \citep{ks15a} in the rendering of the ionizing background. 

Photoionization can be amplified by the presence of a local radiation field from an AGN or an actively star forming galaxy \citep[e.g.,][]{suresh17}. The SDSS distribution of galaxies along the line of sight (see Sec 5) is complete only for galaxies $L > 5L*$. There could be actively star-forming galaxies or relatively faint AGNs in close proximity to the absorber, below the magnitude limit of SDSS spectroscopic survey. Also, as \citet{segers17} and \citet{oppenheimer17} have proposed, galaxies with past AGN activity in time scales shorter than the recombination time scales of high ions ($t \lesssim 10^6$~yrs) can have enhanced {\OVI} ionization fractions in their haloes. We cannot dismiss such a scenario. The size of the proximity zone for a $z=0.2$ AGN with typical $L^*$ luminosity is estimated to be $\sim$ 410 kpc, obtained by taking $L^*$ from \citet{schulze09} with a typical QSO SED from \citet{stevans14} and {\HI} photoionization rate from \citet{gaikwad17a}. One may naively assume that the $4L^*$ galaxy at $850$~kpc projected separation from the absorber (see Table 3) could have hosted a bright AGN of $4L^*$ luminosity in the past. The proximity zone created by such an AGN would be of radius $820$~kpc.  If the AGN was active for more that $2.7$~Myrs ($820$ kpc/$c$), detectable amounts of {\OVI} can still be present in its proximity fossil zone. Along similar lines, we cannot reject the possibility of {\OVI} arising in the proximity fossil zone of some other undetected faint galaxy near to the absorber. However, with the various unknown parameters linked to any past AGN activity, such as the luminosity it had, its spectral energy distribution at {\OVI} ionization energies, the AGN's lifetime or its duty cycle we cannot generate an appropriate local radiation field to include in our ionization models. Hence we restrict our models to the scenario in which photoionization from the extragalactic UV background is dominating the ionization in the gas over any local extreme UV source. Relative heavy element abundances we use in the models are based on the solar reference standard of [C/H]$_{\odot} = -3.57$, [N/H]$_{\odot} = -4.17$, [O/H]$_{\odot} = -3.31$, and [Si/H]$_{\odot} = -4.49$ given by \citet{asplund09}. 

\subsection{Photoionization Models}

Photoionization (PI) calculations were performed using Cloudy (ver. C13.05), last described by \citet{Ferland13}. Cloudy models the absorbing cloud as uniform density plane-parallel slabs in the presence of a radiation field. For the {\HI} column density of the strong saturated core absorption component, we ran a suite of PI models for densities in the range of $-4.0 \leq \log~[n_{\H},~{\cc}] \leq -1.5$, the results of which are shown in Figure 6. 

The density solution for the photoionized gas phase comes from the observed column density ratio between adjacent ionization stages of the same element. From the PI models, we find that the $\log~[N(\CII)/N(\CIII)] = 0.28~{\pm}~0.10$ is valid for $n_{\H} \sim 15\times 10^{-3}$~{\cc}, and $\log~[N(\SiII)/N(\SiIII)] = -0.10~{\pm}~0.11$ is valid for $n_{\H} \sim 3 \times 10^{-3}$~{\cc}, which differ by a factor of five, but comparable within the simplistic assumptions inherent to Cloudy models and the errors in the column density estimates. Both {\CIII} and {\SiIII} have uncertainty in their column density and therefore the estimated densities can be lower, but not too far from these values. 

\begin{figure*} 
\centerline{
\vbox{
\centerline{\hbox{ 
\includegraphics[angle=90,width=0.8\textwidth]{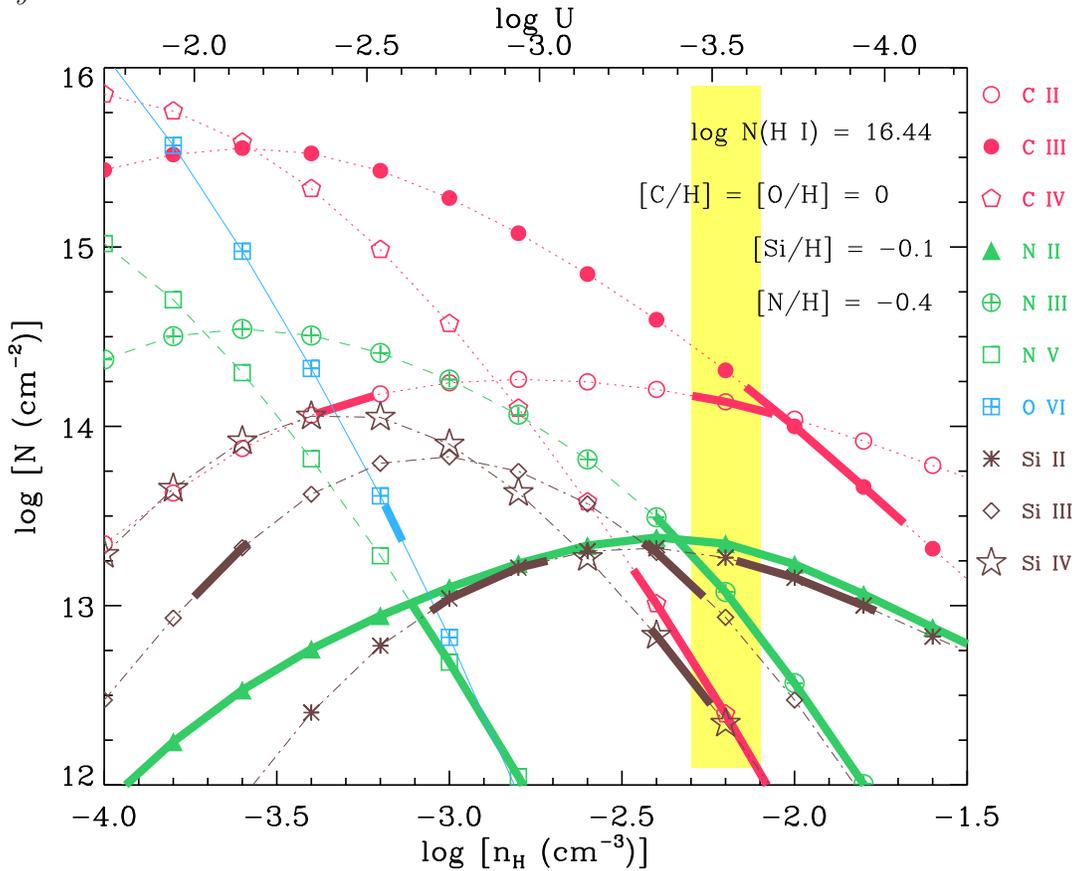} 
}}
}}  
\caption{Photoionization modeling result for the strong {\HI} component. The curves represent the variation of predicted column density with density for the different ions. The observed column density, with its $1\sigma$ uncertainty, is indicated by the thick portion of each curve. The ions of nitrogen are all non-detections and hence provide only an upper limit on the column density. For the given relative elemental abundances, the photoionization models reproduce the observed column densities of {\CII}, {\CIII}, {\SiII}, {\SiIII} and {\SiIV} from the same gas phase with $n_{\H} = (5 - 8) \times 10^{-3}$~{\cc}, indicated by the \textit{yellow} shaded box. This gas phase however produces an {\OVI} column density $6$ dex smaller than observed.}  
\label{fig:5} 
\end{figure*} 

The carbon abundance is constrained by the observed $N(\CII)$. For [C/H] $< -0.1$, the observed $N(\CII)$ is never reproduced by the models for any density. Similarly, {\SiII} is also underproduced at all densities if [Si/H] $< -0.2$. For [C/H] $= 0$ and [Si/H] $= -0.1$, we find the column density ratios between the low and intermediate ionization stages of carbon and silicon agreeing for a narrow range of densities between $n_{\H} = (5 - 8) \times 10^{-3}$~{\cc} (represented by the shaded region of Figure 6). This phase is also consistent with the observed column density of {\SiIV} and the non-detection of {\CIV}. At this density, the non-detection of {\NII}, {\NIII}, and {\NV} require [N/H] $\leq -0.4$. The lower limit of near-solar metallicity for the low ionization material in the absorber is typical of weak {\MgII} absorbers whose metallicities are found to be, on average, an order of magnitude higher than damped {\Lya} absorbers (DLAs) and sub-DLAs \citep{misawa08}. 

Assuming the density of this low ionization phase of the gas to be $n_{\H} = 6 \times 10^{-3}$~{\cc} ($\log~n_{\H} = -2.2$), the PI models yield a total hydrogen column density of $N(\H) = 1.3 \times 10^{18}$~{\cmsq}, electron temperature of $T = 8,550$~K and an absorber path length of $L = 67$~pc. The temperature predicted by the model indicates that the core {\HI} component and metal lines in the photoionized gas are broad due to turbulence, which is consistent with their similar $b$ values. The different $b$-values of {\SiII} (or {\CII}) and {\HI} solve for a temperature of $T \sim 1.9 \times 10^4$~K, with a non-thermal contribution of $b_{nt} \sim 17$~{\kms} to the line widths. The range of temperature predicted by the photoionization models for $n_{\H} = (5 - 8) \times 10^{-3}$~{\cc} is comparable to this. 

Interestingly, the predicted {\OVI} column density from this phase is more than 6 dex lower than its observed value, for solar abundance. The chances of {\OVI} coming from a higher photoionized phase can be ruled out as such a phase would also produce substantial amounts of {\CII}, {\CIII}, {\CIV}, {\SiIII}, and {\SiIV}, which would make the model predictions inconsistent with what the observations indicate. This is evident from Figure 6. The PI modeling clearly suggests that the {\OVI} in this absorber is not a tracer of photonized gas. 

We also considered the possibility of the absorber having a two phase photoionization structure with a low ionization phase tracing {\CII}, {\SiII} and a higher ionization gas responsible for the intermediate ionization {\CIII}, {\SiIII} and the {\SiIV}. Such a model has several limitations. Firstly, a constraint on density, independent of gas metallicity is usually obtained by considering that the adjacent ionization stages of an element (such as {\CII} and {\CIII}, or {\SiII} and {\SiIII}) are tracing the same phase. If instead we assume that the {\CII} and {\CIII} are tracing separate gas phases, we have no unique solution for the density in the absorber. Secondly, the data provides no means to ascertain the {\HI} associated with such a mixture of low and intermediate ionization phase structure. The metallicity that we arrive at, in such a case, for the separate gas phases would at best be only lower limits. Lastly, for the separate intermediate ionization phase to not overproduce {\CII}, the density has to be $\log~n_{\H} < -2.0$ (see Figure 6) such that $N(\CII) < N(\CIII)$. However, this phase traced by {\CIII} would also start contributing {\CIV} significantly, such that $N(\CIV) > N(\CIII)$ for $\log~n_{\H} < -3.5$. However, {\CIV} is a non-detection in the absorber. These limitations compel us to confine to a single phase solution described earlier that explain the low and intermediate ions simultaneously. Irrespective of all these, the PI models clearly suggest that the {\OVI} has to have a separate origin unless the relative elemental abundances are significantly deviant from solar values.    

\subsection{Collisional Ionization Models}

We also considered the collisional ionization equilibrium (CIE) models of \citet{gnat07} to investigate the origin of {\OVI}. The strong {\HI} absorption component is most likely associated with the photoionized, low ionization, high density gas phase explained in the previous section. For the {\OVI} phase, we adopt the {\HI} column density of the BLA component. Figure 7 shows CIE models for a range of equilibrium temperatures. The separate widths of the BLA and {\OVI} set the temperature in this gas phase to $T = (1.8 - 5.2) \times 10^5$~K [$\log~(T/K) = 5.26 - 5.72$]. The models recover the observed column density of {\OVI} from the BLA gas cloud for [O/H] = $-2.5$~dex, at $T = 3.2  \times 10^5$~K, which is very close to the temperature at which the {\OVI} ionization fraction peaks [$f_{\OVI} = N(\mathrm{O^{5+}}/N(\mathrm{O}) \sim 20$\%)] during CIE. At this temperature, the model predicts a total hydrogen column density of $N(\H) = 8.2 \times 10^{19}$~{\cmsq} for the BLA - {\OVI} gas phase, which is $\sim 1.8$ orders of magnitude more than the total hydrogen column density in the photoionized strong {\HI} absorber. The metallicity estimate carries an uncertainty of $\sim~{\pm} 0.17$~dex, imposed by the BLA column density error. The low metallicity is interesting as it is in the range expected for the warm gas in galaxy halos and IGM.  

\begin{figure*} 
\centerline{
\vbox{
\centerline{\hbox{ 
\includegraphics[angle=90,width=1.0\textwidth]{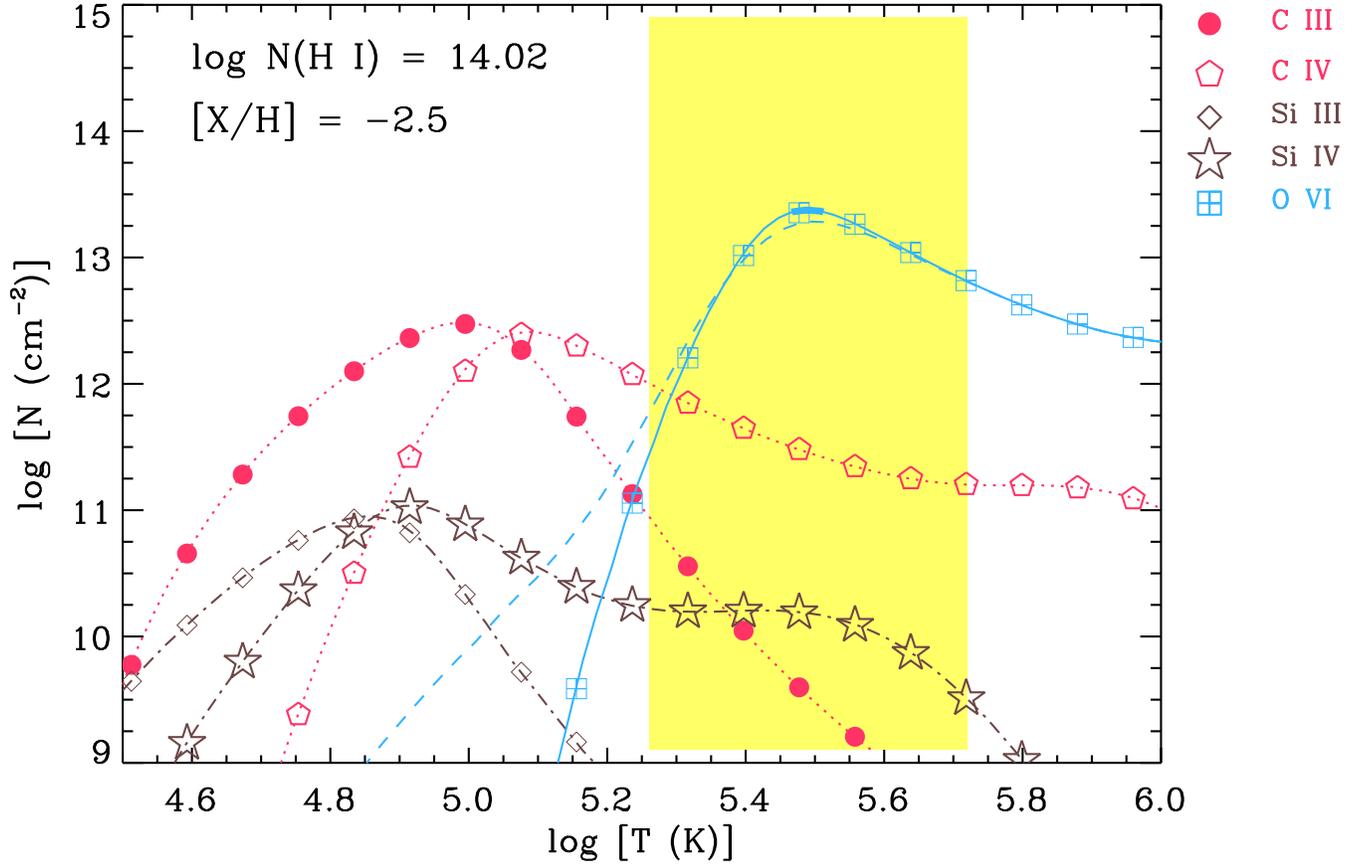} 
}}
}}  
\caption{Column density predictions from CIE models for the intermediate and high ions. The horizontal axis is the collisional ionization equilibrium temperature. The models were generated for the {\HI} column density of the BLA component. The observed {\OVI} is reproduced by the CIE models for [O/H] $= -2.5$~dex, at $T \sim 3 \times 10^5$~K, within the range of temperatures allowed by the $b$ values of the BLA and {\OVI} (\textit{yellow shaded region}). This CIE phase makes insignificiant contributions towards the other ions. The \textit{dashed} curve is the prediction from a constant volume radiatively cooling gas which is not in ionization equilibrium. The curve is for [O/H] $= -1.0$~dex implying that under non-CIE conditions, the abundance in the warm gas phase has to be higher by $\gtrsim 1$~dex than for CIE.}  
\label{fig:6} 
\end{figure*} 

The temperature suggested by the $b(\mathrm{BLA})$ and $b(\OVI)$ is within the $T = (1.8 - 5.2) \times 10^5$~K range where the gas can promptly cool, leading to non-equilibrium ionization fractions \citep[e.g.,][]{savage14}. Under non-CIE conditions, recombination will be delayed compared to the rapidly decreasing temperature of the gas. In our models, this can impact the inferred [O/H] as the ionization fraction of {\OVI} could be lower compared to CIE conditions. The non-CIE calculations of \citet{gnat07} show that the ionization fraction of {\OVI} is lower by $\sim 25$\% at $T = 3.2 \times 10^5$~K and [O/H] $= -2.5$ compared to CIE. This would require the [O/H] in the warm gas phase to be higher by approximately the same factor to match with the observed $N(\OVI)$. The non-CIE predictions are also shown in Figure 7. However, non-equilibrium conditions do not alter the ionization fraction of {\HI} in any appreciable manner. Thus, the total hydrogen column density estimate remains unchanged from the CIE estimate. 

\subsection{A Hybrid Model of Photoionization \& Collisional Ionization Equilibrium}

A more involved collisional ionization model should also include the additional ionization due to extragalactic UV background photons. Ideally such hybrid models should apply the non-CIE scenario, since the estimated gas temperature in our case coincides with the peak of the cooling curve for shock heated gas \citep{gnat07}. However, integrating non-CIE calculations with photoionization is a highly involved process and beyond the scope of the current work. Therefore we restrict the hybrid models to the simultaneous treatment of CIE and photoionization. At the mean temperature of $T = 3.2 \times 10^5$~K set by the $b$-values of BLA and {\OVI} and [O/H] = $-2.5$, the hybrid models predict densities of $n_{\H} \lesssim 10^{-4}$~{\cmsq} for the warm gas phase, with higher oxygen abundances requiring lower densities. This phase has nearly the same {\HI} ionization fraction as CIE, and hence a similar total hydrogen column density of $N(\H) \lesssim 8 \times 10^{19}$~{\cmsq}, and an approximate line of sight thickness of $L \sim 265$~kpc. 

\section{Galaxies Near the Absorber}

The quasar field is covered by SDSS. A search through SDSS DR13 database \citep{SDSS2016} for extended sources within 5 Mpc of projected separation and $\Delta v < 800$~{\kms} of the absorber reveals 12 galaxies. These galaxies listed in Table 3 have an internal velocity dispersion of $\sim 517$~{\kms}, with a mean velocity of $45$~{\kms} relative to the absorber. The nearest galaxy is at 858 kpc distance. We estimate the halo radius of this galaxy to be $R_{vir} = 287$~kpc using the scaling relationship between $R_{vir}$ and $L$ given by \citet{stocke14}. The galaxy is thus too far out in projected separation to be causally linked to the absorber. 

\begin{figure*}
\begin{subfigure}{0.35\textwidth}
\includegraphics[angle=90,scale=0.5]{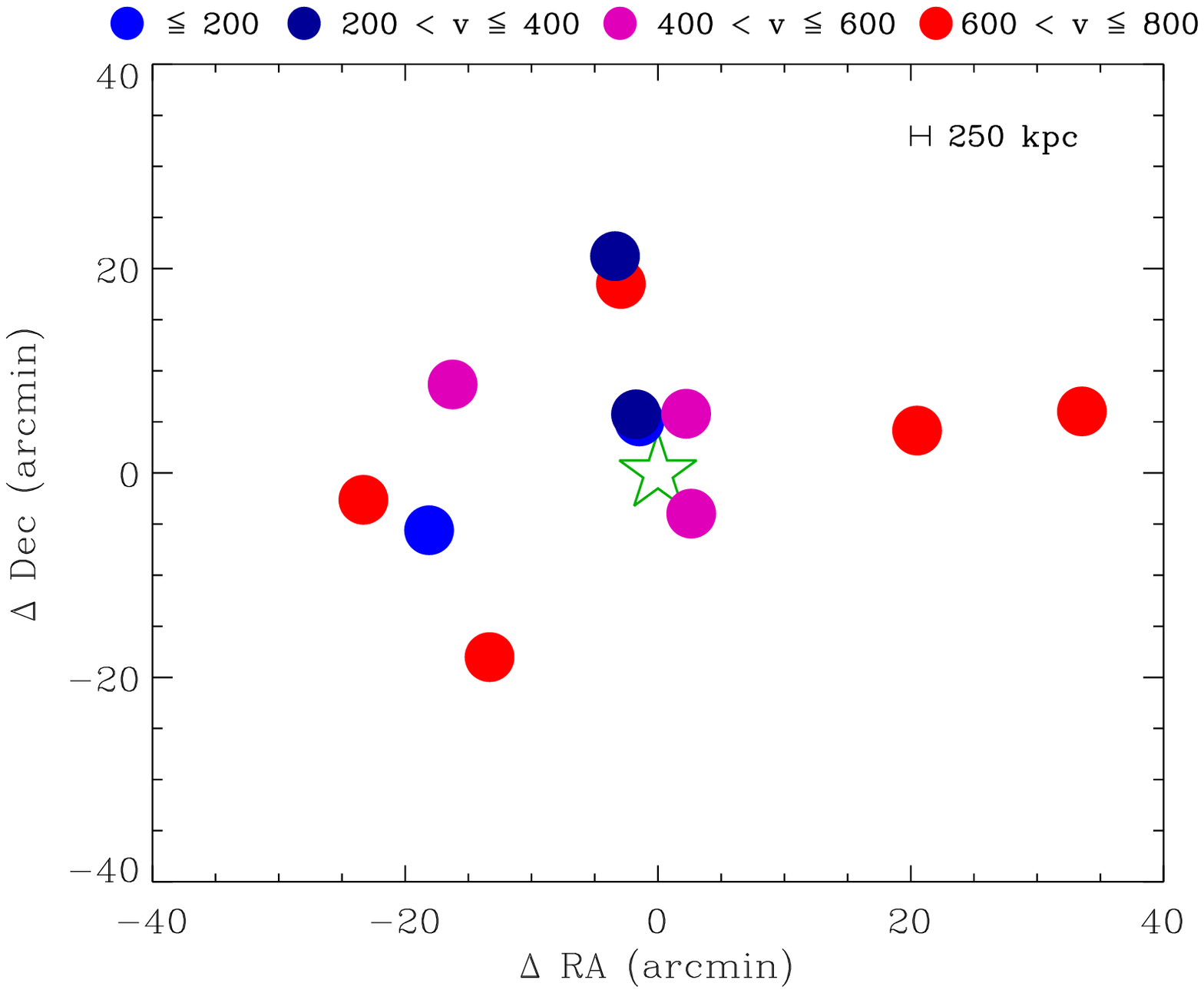} 
\end{subfigure}\hspace{0.15\textwidth}
\begin{subfigure}{0.35\textwidth}
\includegraphics[angle=90,scale=0.48]{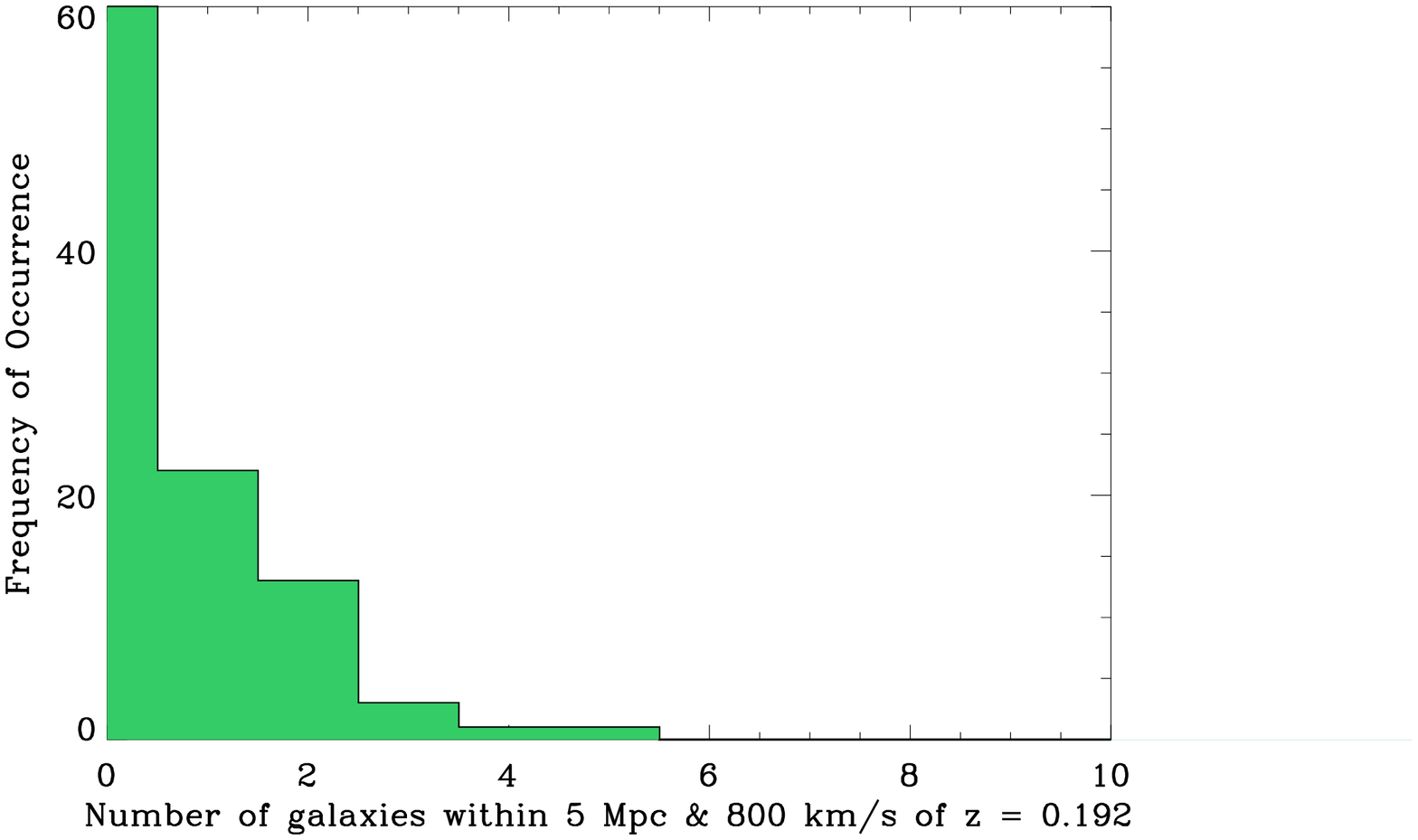}
\end{subfigure} 
\caption{The \textit{left panel} shows the distribution of the 12 SDSS galaxies listed in Table~\ref{tab:tab3} relative to the PG~$1121+422$ quasar sightline. The quasar position is indicated by the \textit{star} symbol at ($0,0$) coordinates. The galaxies are the \textit{filled circles} where the colors indicate the velocity offset in {\kms} of the absorber from the systemic velocity of each galaxy. For reference, the physical scale of $250$~kpc, corresponding to the approximate virial radius of an $L^*$ galaxy, is also indicated. The \textit{right panel} is the frequency distribution of finding a certain number of galaxies in the SDSS database by random coincidence. The distribution was obtained by sampling a $5$~Mpc, $|\Delta v| = 800$~{\kms} search window around a 100 random locations at $z=0.19$. The probability of finding 6 or more galaxies by random coincidence is $< 1$\%}  
\label{fig:6} 
\end{figure*}

The SDSS spectroscopic data is $90$\% complete down to an approximate $r$-band magnitude of $r < 17.8$, equal to a luminosity of $\gtrsim 5~L*$ at $z \sim 0.2$ \citep{ilbert05}. Thus, SDSS is sampling only the most luminous galaxies at the redshift of this absorber. To verify whether the incidence of so many luminous galaxies within our chosen search window is unusual, we sampled $100$ random regions of SDSS for galaxies at $z=0.19$ within a similar search window. The results of the sampling, shown in Figure 9, indicates a less than 1\% probability of finding 6 or more luminous galaxies in a similar region of space by random coincidence. The absorber thus seems to be residing in a galaxy overdensity region, which is consistent with its physical properties, as warm {\HI} gas is known to be strongly correlated with the spatial distribution of galaxies and large-scale filaments \citep{stocke14, nevalainen15, wakker15, pachat16, tejos16}.

\begin{table*}
\centering
\caption{Galaxies Proximate to the Absorber}
\scriptsize
\begin{tabular}{c c c c c c c c c c}
\hline
\hline
RA (J2000)     &   Dec (J2000)  &  $z_{gal}$   &  $\Delta v$~(\kms)  &   $\eta$~(arcmin)  &  $\rho$~(Mpc) &  g (mag)   &  r (mag)  & M$_g$ (mag)  & ($L/L^*$)$_g$ \\[2ex] 
\multicolumn{9}{c}{SDSS Galaxies} \\ \hline
171.11949  &    42.09552 &    $0.19399~{\pm}~0.00004$ &  $403~{\pm}~11$  &    4.4325 &    0.8580 &   $18.9$ &   $17.6$ &    $-21.7$ & $4.1$    \\
171.18777  &    41.94499 &    $0.19222~{\pm}~0.00004$ &  $-42~{\pm}~11$  &    5.1683 &    1.0004 &   $18.5$ &   $17.2$ &    $-22.0$ & $5.6$    \\
171.19225  &    41.93347 &    $0.19151~{\pm}~0.00005$ &  $-222~{\pm}~13$ &    5.8865 &    1.1394 &   $18.6$ &   $17.5$ &    $-21.8$    & $4.4$    \\
171.12613  &    41.93261 &    $0.19432~{\pm}~0.00004$ &  $485~{\pm}~11$  &    6.0266 &    1.1665 &   $18.2$ &   $16.8$ &    $-22.4$ & $7.9$    \\
171.46507  &    42.12255 &    $0.19202~{\pm}~0.00004$ &  $-93~{\pm}~11$  &   14.5623 &    2.8187 &   $19.0$ &   $17.7$ &    $-21.5$    & $3.5$    \\
171.43398  &    41.88480 &    $0.19035~{\pm}~0.00004$ &  $-512~{\pm}~11$ &   14.8642 &    2.8771 &   $18.9$ &   $17.6$ &    $-21.6$    & $3.7$    \\
170.82166  &    41.95994 &    $0.19540~{\pm}~0.00004$ &  $755~{\pm}~11$  &   15.7900 &    3.0563 &   $18.3$ &   $17.1$ &    $-22.2$    & $6.6$    \\
171.55173  &    42.07291 &    $0.18987~{\pm}~0.00004$ &  $-635~{\pm}~11$ &   17.5048 &    3.3882 &   $17.9$ &   $16.5$ &    $-22.7$    & $9.8$    \\
171.21216  &    41.72097 &    $0.18959~{\pm}~0.00004$ &  $-704~{\pm}~11$ &   18.6207 &    3.6042 &   $18.5$ &   $17.2$ &    $-21.9$    & $5.3$    \\
171.38535  &    42.32946 &    $0.19518~{\pm}~0.00003$ &  $701~{\pm}~8$   &   20.5452 &    3.9767 &   $18.5$ &   $17.2$ &    $-22.1$    & $5.8$    \\
171.21986  &    41.67559 &    $0.19124~{\pm}~0.00005$ &  $-290~{\pm}~13$ &   21.3648 &    4.1354 &   $17.9$ &   $16.5$ &    $-22.7$    & $9.9$    \\
170.60429  &    41.92870 &    $0.19514~{\pm}~0.00004$ &  $691~{\pm}~11$  &   25.6506 &    4.9649 &   $18.8$ &   $17.6$ &    $-21.7$    & $4.1$    \\
\hline
\hline
\centering
RA (J2000)     &   Dec (J2000)  &  $\eta$~(arcmin)  &  $\rho$~(Mpc) &  $m_{\mathrm{AB}}$ (mag)  & Label \\[2ex] 
\multicolumn{6}{c}{NICMOS galaxies} \\ \hline
$171.1372321$ & $42.0310633$  & $0.1973$ & $0.0378$ & $22.3$ & N2 \\
$171.1372759$ & $42.0337335$  & $0.2293$ & $0.0439$ & $22.2$ & N3 \\
$171.1366955$ & $42.0262385$  & $0.3911$ & $0.0749$ & $21.8$ & N1 \\
$171.1318351$ & $42.0274755$  & $0.5017$ & $0.0961$ & $19.5$ & N4 \\
$171.1253927$ & $42.0299198$  & $0.7292$ & $0.1397$ & $22.2$ & N5 \\
\hline
\hline
\end{tabular}
\label{tab:tab3}
\scriptsize{Comments - In the \textit{top} table are the galaxies as seen by SDSS within $5$~Mpc of projected separation from the line of sight and $|\Delta v| = 800$~{\kms} of the absorber. Column 3 lists the spectroscopic redshifts of the galaxies, column 4 is the systemic velocity of the galaxy relative to the absorber, column 5 is the projected separation in the plane of the sky between the line of sight to the background quasar and the location of the galaxy, column 6 is the projected separation in Mpc (derived using an angular scale of $3.226$ kpc (arcsec)$^{-1}$ at $z=0.19$, \citet{wright06}), columns 7 \& 8 are the apparent magnitudes in the SDSS $g$, and $r$ bands, columns 9 \& 10 list the absolute magnitudes and the luminosities of the galaxies after including K-correction. The K-corrections were calculated using the analytical expression given by \citet{chilingarian10}. The galaxy absolute magnitudes were calculated by determining the luminosity distances for a flat $\Lambda$CDM universe with $H_0 = 69.6$~{\kms}~Mpc$^{-1}$, $\Omega_m = 0.286$, $\Omega_{\Lambda} = 0.714$ \citep{bennett14}. The Schechter absolute magnitude $M^*_g = -20.2$ for $z = 0.19$ was taken from \citet{ilbert05}, and carries an uncertainty of $\sim 0.2$~mag, which translates into a luminosity uncertainty of $\sim 20$\%. The \textit{bottom} segment of the table lists galaxies identified in the single band NICMOS image of the quasar field. The AB magnitudes listed for NICMOS galaxies are isophotal magnitudes in the F160W filter.}
\end{table*}

The $N(\HI) \sim 10^{16.4}$~{\cmsq} of this absorber is reminiscent of circumgalactic gas. Absorber-galaxy surveys show correlations of high column densities (and strong equivalent widths) with regions that are not too far from galaxies \citep[$\rho \lesssim 300$~kpc, $|\Delta v| \lesssim 400$~{\kms}][]{wakker09,prochaska17}. If such a galaxy exists for our absorber, it will have to be of $\lesssim L^*$ luminosity. The galaxy will also have to be significantly closer ($\rho \lesssim 200$~kpc) to the line-of-sight since the size of gas haloes is known to scale with galaxy luminosity \citep{chen98,chen01b,kacprzak08}. As SDSS is sampling only the bright end of the galaxy luminosity function, it is within reason to assume that there are many more galaxies fainter than $L^*$ within the same field of view. Single band NICMOS/F160W images, of comparatively smaller field of view ($0.9 \times 0.9$~sq.arcmin), for this field show 5 additional galaxies fainter than $19$ AB-magnitude within 0.75 arcminute ($\rho < 150$~kpc) of the absorber whose redshifts are unknown (see Table 3). If any of these NICMOS galaxies are coincident in redshift with the absorber, it would place the absorber within their virial radii. With the data available to us, it is difficult to establish whether the warm absorber is circumgalactic material associated with an $\lesssim L^*$ galaxy or intergalactic gas within a group environment. Deeper galaxy observations yielding redshift information is required to explore these scenarios further. 

\section{Summary \& Discussion}

The main results of our analysis are as follows:

\begin{enumerate}

\item The absorber at $z = 0.19236$ has {\HI}, {\CII}, {\SiII}, {\CIII}, {\SiIII}, {\SiIV} and {\OVI} lines detected. The rest-frame equivalent widths of {\CII} and {\SiII} indicate that this is a weak {\MgII} class of absorber. The {\OVI} associated with the absorber is also weak.   

\item The {\HI} and metal lines have kinematically simple profiles. The metal lines are all adequately explained by a single component, whereas the {\Lya} shows the need for a broad component with $b(\HI) = 71^{+22}_{-14}$~{\kms} to explain the absorption along its wings. The simple kinematic nature of this absorber, with two or less absorbing components, is typical of the {\OVI} absorber population detected at low redshifts. 

\item Ionization models suggest the presence of multiple gas phases in the absorber. The low ionization gas traced by {\CII}, {\SiII}, {\CIII}, {\SiIII} and {\SiIV} is a predominantly photoionized medium with $n_{\H} \sim 6 \times 10^{-3}$~{\cc}, $T \sim 8,550$~K, total hydrogen column density of $N(\H) \sim 1.3 \times 10^{18}$~{\cmsq} and absorber line of sight thickness of $L \sim 65$~pc, with [C/H] $\sim 0$, [Si/H] $\sim -0.2$~dex, and [N/H] $\leq -0.7$~dex, with an uncertainty of ${\pm}~0.1$~dex in the metallicity. The column density of {\OVI} from this photoionized gas is $\sim 6$~dex smaller than its observed value.

\item The {\Lya} is best fitted with a three component model where the weaker component is a BLA with $\log~N(\HI) = 14.02~{\pm}~0.17$ and $b(\H) = 71^{+22}_{-14}$~{\kms}. The different $b$-values of {\OVI} and BLA solve for a temperature of $T = (1.8 - 5.2) \times 10^5$~K.

\item In CIE models, the weak {\OVI} is produced in low metallicity gas ([O/H] = $-2.5~{\pm}~0.17$) at $T = 3.2 \times 10^5$~K, and $N(\H) = 8.2 \times 10^{19}$~{\cmsq}, which is $\sim 1.8$ orders of magnitude greater than the column density of total hydrogen in the cooler photoionized medium of the absorber. Models that integrate both photoionization and CIE calculations predict the density in this warm gas phase to be $n_{\H} \lesssim 10^{-4}$~{\cc}. 

\item The SDSS spectroscopic database shows 12 galaxies within $\rho = 5$~Mpc and $|\Delta v| = 800$~{\kms} of the absorber suggesting that the absorption is tracing a galaxy overdensity region, with the nearest galaxy at a projected separation of $\rho = 858$~kpc. A NICMOS single band image of the quasar field reveals galaxies within several arcseconds of the sightline, much fainter than the detection threshold of SDSS. The absorber could be well within the virial radii of these galaxies if they are coincident with the absorber. Establishing the redshift of these galaxies through follow-up ground based observations can lead to a conclusion on whether this warm absorber is tracing multiphase intragroup gas or the CGM of a $\lesssim L^*$ galaxy. 

\end{enumerate}

Building on the important low redshift {\OVI} survey with $HST$/STIS of \citet{tripp08}, \citet{savage14} found in a much higher $S/N$ blind survey with COS of 54 {\OVI} absorption systems that in as many as half the number of cases, the absorption systems are kinematically simple with a single component for {\OVI} and a superposition of a narrow and broad component for {\HI}. The absorber described in this paper shares this kinematic simplicity. The BLA and {\OVI} integrated column densities are comparable to their respective median values from the larger sample of \citet{savage14}. The low ionization lines also possess a similar kinematic structure. This strikingly repetitive trend could be hinting at a significant fraction of warm {\OVI} absorbers tracing a simple two-phase temperature-density medium, such as in the case of highly ionized Galactic high velocity clouds (HVCs). 

The {\OVI} in Galactic HVCs are produced through collisional ionization in transition temperature gas formed at the interface layers between a dense $n_{\H} \sim (1 - 10) \times 10^{-3}$~{\cc} photoionized $T \sim 10^4$~K cloud and a $T \sim 10^6$~K coronal halo in which the cloud is embedded \citep{sembach03, fox04, fox05, wakker12}. The near alignment of the cold and warm gas phases, the $T > 10^5$~K temperature and the relatively low {\OVI} column density of $N(\OVI) = 10^{13.4}$~{\cmsq} for the absorber agrees with the general predictions from interface models \citep{boehringer87,savage06}. If the line of sight is intercepting a galaxy group environment, as suggested by the excess of bright  galaxies at the location of the absorber, the {\OVI} can also form at the interface between the diffuse and hot ($T \sim 10^{6.5}$~K) intra-group medium and the $T \lesssim 10^4$~K photoionized clouds within the group \citep{stocke14, stocke17}. \citet{bielby17} provide an example where such low ionization gas clouds are relics of past outflows from galaxies that belong to the group. 

This view of collisionally ionized interface gas is also compatible with predictions of cosmological simulations that model accretions and outflows from galaxies. Such simulations show {\CII}, {\SiII}, and {\MgII} tracing compact halo clouds belonging to galactic outflows of recent origin, with the high-ions such as {\OVI} coming from a more extended and diffuse circumgalactic medium shaped by ancient outflows older than a billion years \citep{oppenheimer12, ford14, ford16}. Such a scheme also explains the higher proportion of metals we find in the absorber's low ionization gas compared to the {\OVI} phase, and their different thicknesses along the line-of-sight. 

However, the same simulations of \citet{oppenheimer09} and \citet{oppenheimer12} end up predicting average temperatures of $T \sim 1.5 \times 10^4$~K for the {\OVI}, which does not accord well with the absorber discussed here, or the 14 absorbers in the larger sample of \citet{savage14}. The derived temperatures from the combined BLA - {\OVI} line widths are a better match with the $T \sim 2 \times 10^5$~K value obtained by \citet{tepper11} and \citet{smith11} in their cosmological simulations that employ photoionization and collisional ionization simultaneously to compute gas ionization fractions. Such departures of simulation predictions from observations are expected because of the different assumptions and treatment of physics built into the various simulations, and the wide range of physical properties that the {\OVI} absorbers embody. It reaffirms the need to analyze individual absorption systems in detail, to get close to their true physical and chemical properties, even as large absorption line surveys bring out global statistics on these absorber populations. Specifically, the analysis of this absorber reasserts the importance of {\OVI}-BLA absorbers as direct probes of shock-heated tenuous gas with $T \gtrsim 10^5$~K. High $S/N$ spectroscopic observations using COS are required to discover shallow BLA features and weak {\OVI} in multiphase absorbers, and the low metallicity, warm baryonic gas reserves they trace outside of galaxies.  

\section{Acknowledgments}

The authors wish to thank the people involved in the design, building and maintenance of COS and the HST. AN thanks his host institution and the Department of Space, Government of India for financial support. Wakker acknowledges support from HST program GO-14588 provided by NASA through a grant from the Space Telescope Science Institute, which is operated by the Association of Universities for Research in Astronomy, Inc, under NASA contract NAS5-26555. The study made use of the NASA/IPAC Extragalactic Data Base (NED) which is operated by the Jet Propulsion Laboratory, California Institute of Technology under contract with NASA. 

\def\aj{AJ}%
\def\araa{ARA\&A}%
\def\apj{ApJ}%
\def\apjl{ApJ}%
\def\apjs{ApJS}%
\def\apss{Ap\&SS}%
\def\aap{A\&A}%
\def\aapr{A\&A~Rev.}%
\def\aaps{A\&AS}%
\def\mnras{MNRAS}%
\def\memras{MmRAS}%
\def\pasp{PASP}%
\def\nat{Nature}%
\def\aplett{Astrophys.~Lett.}%
\def\procspie{Proc.~SPIE}%
\let\astap=\aap
\let\apjlett=\apjl
\let\apjsupp=\apjs
\let\applopt=\ao
\bibliographystyle{mn}
\bibliography{mybib}

\end{document}